\documentclass[a4paper,fleqn]{cas-dc}
\usepackage[numbers]{natbib}
\usepackage{float}
\usepackage{svg} 
\usepackage{soul}
\usepackage{xcolor}
\sethlcolor{yellow}

\def\tsc#1{\csdef{#1}{\textsc{\lowercase{#1}}\xspace}}
\tsc{WGM}
\tsc{QE}
\tsc{EP}
\tsc{PMS}
\tsc{BEC}
\tsc{DE}
\begin{document}

\let\WriteBookmarks\relax
\def\floatpagepagefraction{1}
\def\textpagefraction{.001}
\shorttitle{CNN and AST for Sound Classification}    
\shortauthors{P.B. Dehaghani et~al.}

\author[1]{Parinaz Binandeh Dehaghani}
\ead{up202100618@edu.fe.up.pt}
\cormark[1]
\author[2]{Danilo Pena}
\ead{danilo.pena@resosight.com}
\author[1]{A. Pedro Aguiar}
\ead{pedro.aguiar@fe.up.pt}

% -----------------------
% Affiliations
% -----------------------

\address[1]{SYSTEC-ARISE, Faculty of Engineering University of Porto, Portugal}
\address[2]{ResoSight, Montreal, Canada}

%\title [mode = title]{}  
\title [mode = title] {Evaluating CNN with Stacked Feature Representations and Audio Spectrogram Transformer Models for Sound Classification}

\begin{abstract}
Environmental sound classification (ESC) has gained significant attention due to its diverse applications in smart city monitoring, fault detection, acoustic surveillance, and manufacturing quality
control. To enhance CNN performance, feature stacking techniques have been explored to aggregate complementary acoustic descriptors into richer input representations. In this paper, we investigate
CNN-based models employing various stacked feature combinations, including Log-Mel Spectrogram (LM), Spectral Contrast (SPC), Chroma (CH), Tonnetz (TZ), Mel-Frequency Cepstral Coefficients (MFCCs), and Gammatone Cepstral Coefficients (GTCC). Experiments are conducted on the widely used ESC-50 and UrbanSound8K datasets under different training regimes, including pretraining on ESC-50, fine-tuning on UrbanSound8K, and comparison with Audio Spectrogram Transformer (AST) models pretrained on large-scale corpora such as AudioSet. This experimental design enables an analysis of how feature-stacked CNNs compare with transformer-based models under varying levels of training data and pretraining diversity. The results indicate that feature-stacked CNNs offer a more computationally and data-efficient alternative when large-scale pretraining or extensive training data are unavailable, making them particularly well suited for resource-constrained and edge-level sound classification scenarios.
\end{abstract}


% Research highlights%
%\begin{highlights}    \item Feature-stacked CNN architectures achieve strong validation performance (up to 0.92) when trained with moderate transfer (ESC → US8K), demonstrating effective feature utilization under constrained data conditions. \item Both CNN and AST models benefit from cross-dataset transfer. However, stacked CNNs maintain competitive performance with relatively lower computational complexity compared to Transformer-based models.\item When large-scale pretraining is available (AudioSet), the AST model achieves the highest validation accuracy (0.99), highlighting the impact of large-scale representation learning rather than purely architectural superiority. \item The results emphasize that data scale and pretraining strategy strongly influence performance, and model selection should consider the available transfer-learning regime alongside computational constraints.\end{highlights}

% Keywords

\maketitle
% Main text
\section{Introduction}
Environmental sound classification (ESC) has emerged as a crucial task in machine learning and signal processing,
finding applications in diverse domains such as surveillance, smart city monitoring, fault detection in automobiles,
and acoustic-based quality control in manufacturing. Unlike structured audio signals such as speech and music,
environmental sounds exhibit significant variations in temporal and spectral characteristics, making classification a
challenging problem.
Recent advances in machine learning and deep learning have transformed the way environmental audio is analyzed ~\cite{Purwins2019}.
First, traditional machine learning approaches relied on handcrafted features such as Mel- Frequency Cepstral Coefficients
(MFCCs), Zero-Crossing Rate (ZCR), and Spectral Contrast (SPC), among others, and their combinations \cite{chachada2014environmental, Bountourakis2015}.
They were used as inputs to classifiers like Support Vector Machines (SVMs), k-Nearest Neighbors (k-NN), Decision
Trees (DT), Local Binary Pattern (LPB), and Random Forest (RF) ~\cite{Roma2018, Toffa2021}. These models achieved reasonable results but
often lack robustness to noise and generalization across diverse acoustic environments. 
With the advent of deep learning, particularly Convolutional Neural Networks (CNNs), ESC systems became capable of learning hierarchical time-frequency patterns directly from the spectrogram representations ~\cite{Purwins2019}. They can efficiently extract local spectrum features, making them well-suited for multiple audio classification tasks ~\cite{hershey2017cnn, Zheng2018, Ghosal2018}.  

However, CNNs often struggle with capturing long-range dependencies in audio sequences, which is a fundamental limitation in modeling complex environmental sounds. In contrast, transformer-based models, such as Audio Spectrogram Transformers (AST)~\cite{gong2021ast}, leverage self-attention mechanisms to model global dependencies in spectrogram representations. ASTs have demonstrated state-of-the-art performance in various audio classification tasks, particularly when pre-trained on large-scale datasets ~\cite{gong2021ast}. However, their reliance on extensive data and computational resources poses challenges for real-world applications where labeled data is limited. Within this broader evolution, researchers have also explored feature aggregation as a complementary approach to improve learning efficiency ~\cite{Liu2021}. By combining multiple extracted features, researchers have sought to improve classification performance and reduce model complexity, making systems more adaptable to various environmental conditions.
While CNNs are commonly employed as standalone feature extractors, studies investigating their enhancement through feature aggregation highlight the potential for improved generalization and robustness~\cite{Su2020}. This approach has been explored in both shallow machine learning models~\cite{Domazetovska2021} and deep learning frameworks~\cite{Liu2021}, emphasizing its role in refining ESC methodologies. 

This study presents a performance comparison focused primarily on CNN models utilizing stacked feature representations for ESC using the widely adopted ESC-50 dataset. We investigate the impact of different stacked feature combinations as input to CNN, including Log-Mel Spectrogram (LM), Spectral Contrast (SPC), Chroma (CH), Tonnetz (TZ), and Mel-Frequency Cepstral Coefficients (MFCCs). The evaluated combinations include LM, LM+ TZ, LM+ MFCC,  MFCC+ TZ, LM+ SPC+ CH, and MFCC+ GTCC+ CH+ LM. Additionally, we have evaluated the CNNs against the AST model, which is known for learning long-range dependencies from audio data. The findings of this study provide insights into the effectiveness of stacked feature representations in CNN-based ESC and highlight the trade-offs between CNN and AST architectures for environmental sound classification.

\section{Related Work} 
Feature aggregation, also referred to as feature concatenation or feature-level fusion, allows models to leverage complementary information from different spectral and temporal characteristics of environmental sounds. Studies have investigated various feature aggregation approaches, including the concatenation of Log-Mel Spectrograms, Spectral Contrast, Chroma, Tonnetz, and MFCCs to enhance classification accuracy~\cite{gupta2024esc,  Mushtaq2020_2, sharma2019environment, zhu2018environmental}. Feature selection and dimensionality reduction techniques, such as Principal Component Analysis (PCA), have also been applied to aggregated features to reduce redundancy and computational complexity~\cite{Burgos2014}. 

Deep learning has revolutionized ESC by enabling automatic feature extraction from raw audio signals or spectrogram representations. CNNs have been widely adopted for ESC due to their ability to capture local spectral-temporal features.
They are commonly employed to extract features directly from spectrogram representations, which are then passed to a neural network classifier~\cite{piczak2015environmental}. However, to improve classification performance and develop high-efficiency AI models, some studies have explored feature aggregation as a strategy. These methods involve combining multiple spectral features, allowing models to leverage complementary information from different aspects of the audio signal.
More advanced fusion strategies, such as attention-based fusion~\cite{Dai2020} and multimodal feature integration, have also been proposed to enhance ESC performance~\cite{sharma2019environment}. However, most existing aggregation research focuses on feature aggregation as a technique within CNN models, without directly comparing its impact against standalone CNNs or ASTs. This lack of direct comparison highlights the contribution of our study, which systematically evaluates the performance of CNNs trained with single features versus CNNs with stacked features, including evaluation against the AST model. Studies have explored enhanced CNN architectures, including deep convolutional networks with
regularization techniques and data augmentation to improve robustness~\cite{Mushtaq2020}. Nevertheless, the direct comparison between standalone CNNs and improved CNNs utilizing stacked feature fusion remains underexplored. 

Transformers, particularly ASTs, while not the primary focus of this study, have recently gained attention for ESC due to their ability to model long-range dependencies in audio signals. ASTs utilize self-attention mechanisms to capture contextual information across the entire spectrogram representation. Recent studies have demonstrated the effectiveness of ASTs in ESC, achieving state-of-the-art results on benchmark datasets such as ESC-50 and UrbanSound8K ~\cite{chen2021environmental, salamon2014dataset}.
However, AST models typically require extensive pre-training and large-scale datasets to achieve optimal performance, making them computationally demanding for real-time applications. Thus, larger datasets such as Audioset have been used for pretraining purposes, leveraging the AST performance in ESC applications. In this work, we compare the AST using the same datasets as the CNNs, ESC-50 and UrbanSound8K, also with a larger dataset (Audioset), against the CNN and the CNN with stacked features. Despite these advancements, there is a lack of direct comparison between CNN models with stacked feature representations, including evaluations with other architectures, particularly AST, in the context of ESC. This study aims to address this gap by conducting a comprehensive evaluation of CNNs with stacked features on the UrbanSound8K dataset, analyzing how feature aggregation influences classification performance. Additionally, we provide a comparative analysis against AST models to assess how well CNNs with aggregated features capture spectro-temporal patterns compared to self-attention-driven transformer models. This comparison serves to contextualize the efficiency of CNN-based feature stacking strategies in relation to AST architectures.

\section{Methodology} 
In this section, we describe the methodology employed for environmental sound classification using CNNs. Our approach
involves extracting and stacking multiple feature representations to capture a more comprehensive understanding of environmental sounds, thus enhancing model performance. We first extract key acoustic features from the audio signals. These features are then aggregated into stacked representations, which serve as input to our deep-learning models. We compare the effectiveness of different feature combinations in improving classification accuracy. 

\subsection{Features}\label{Features}
To extract meaningful features from audio signals, we utilized the Librosa library ~\cite{mcfee2015librosa}, which provides multiple methods for feature extraction in environmental sound classification. Extracted features include cepstral-based features, spectral features, and tonal features, which are essential for capturing various acoustic properties of environmental sounds.
Below, we describe each extracted feature in detail. 

\subsubsection{Mel-Frequency Cepstral Coefficients (MFCCs)}
\begin{figure*}
    \centering
    \includegraphics[width=0.8\linewidth]{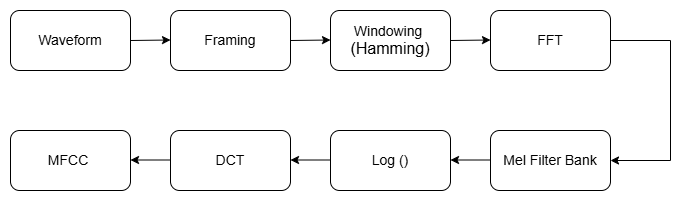}
    \caption{Block diagram of MFCC algorithm}
    \label{fig:Block diagram of MFCC algorithm}
\end{figure*}

MFCCs are derived from the log power spectrum of an audio signal ~\cite{Gourisaria2024}, using the Mel scale, which better reflects human auditory perception compared to a linear frequency scale. As illustrated in Figure\ref{fig:Block diagram of MFCC algorithm}, the computation begins with the raw waveform, which is segmented into short overlapping frames. Each frame is then windowed using a Hamming window to minimize spectral leakage. The windowed frames undergo a Fast Fourier Transform (FFT) to convert the
time-domain signal into the frequency domain. Subsequently, a Mel filter bank is applied to the magnitude spectrum to emphasize perceptually important frequency bands. The filtered spectral energies are converted to a logarithmic scale, and a Discrete Cosine Transform (DCT) is applied to decorrelate the features. The resulting coefficients are the MFCCs, which compactly represent the spectral envelope of the audio signal. The n-th cepstral coefficient, Cn, representing the spectral envelope features, can be represented as: 

\begin{equation}
C_n = \sum_{m=0}^{M-1} \log(S_m)\,
\cos\left[ n\left(m - 0.5\right)\frac{\pi}{M} \right],
\end{equation} 

where $M$ is the total number of filter banks, $S_m$ is the spectral energy  in the $m$-th frequency band (computed from the filter bank output),  $n$ is the index of the cepstral coefficient (ranging from $0$ to $N-1$,  where $N$ is the number of coefficients extracted), 
$m$ is the index of the filter bank (ranging from $0$ to $M-1$). 
$\log(S_m)$ is the logarithm of spectral energy, used to mimic human auditory perception. 

\subsubsection{Log-Mel Scale Spectrogram (LM)} 
\begin{figure*}
    \centering
    \includegraphics[width=0.8\linewidth]{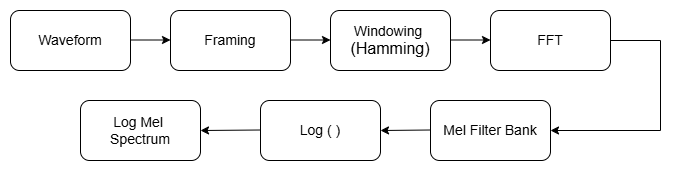}
    \caption{Block diagram of Log-Mel Scale Spectrogram algorithm}
    \label{fig:Block diagram of Log-Mel Scale Spectrogram algorithm}
\end{figure*}

A Log-Mel spectrogram~\cite{Su2020} is a time-frequency representation of an audio signal in which the frequencies are mapped to the Mel scale and then transformed to a logarithmic scale. This representation provides a compact and perceptually relevant view of the spectral content, making it a popular choice for deep learning models. A Log-Mel spectrogram, as
shown in Figure \ref{fig:Block diagram of Log-Mel Scale Spectrogram algorithm}, is computed in a manner similar to MFCCs, but without applying the Discrete Cosine Transform (DCT). The Log-Mel Spectrogram value, Log-Mel(m), for the m-th Mel filter is: 

\begin{equation}
\mathrm{Log\text{-}Mel}(m) 
= \log \left( 
\sum_{k=0}^{N-1} 
\left| X(n,k) \right|^2 H_m(k) + \epsilon 
\right),
\end{equation}

where $N$ is the total number of frequency bins in the spectrogram. 
$X(n,k)$ is the Short-Time Fourier Transform (STFT) coefficient at time frame $n$  and frequency bin $k$. 
$|X(n,k)|^2$ represents the power spectrum (magnitude squared of the STFT). 
$H_m(k)$ is the Mel filter bank response at the $m$-th filter and frequency $k$,  which converts the linear frequency scale to the Mel scale. 
$\epsilon$ is a small constant added for numerical stability to avoid taking the logarithm of zero. 

\subsubsection{Chroma Features (CH)} 
Chroma features~\cite{Su2020} aggregate spectral information into 12 distinct pitch classes. These features emphasize the harmonic and tonal structure of a sound. To compute them, first perform the STFT of the audio signal, then map frequency bins to 12 chroma bins corresponding to pitch classes. The Chroma feature is given by:

\begin{equation}
C_n = 
\frac{\sum_{k \in B_n} \left| X(n,k) \right|^2}
{\sum_{k=0}^{N-1} \left| X(n,k) \right|^2},
\end{equation}

where $B_n$ is the set of frequency bins corresponding to pitch class $n$. 
$|X(n,k)|^2$ is the STFT power spectrum.

\subsubsection{Spectral Contrast (SPC)} 
Spectral contrast~\cite{Su2020} measures the difference between peaks and valleys in the frequency spectrum rather than focusing on the average spectral envelope. It captures tonal richness and helps distinguish between harmonic and percussive sounds. The Spectral Contrast feature can be described as: 

\begin{equation}
SC_m = 
\frac{S_{\max,m} - S_{\min,m}}
{S_{\max,m} + S_{\min,m}},
\end{equation}

where $S_{\max,m}$ is the maximum spectral amplitude in band $m$ and 
$S_{\min,m}$ is the minimum spectral amplitude in band $m$.

\subsubsection{Tonnetz (TZ)} 
Tonnetz (Tonal Centroids) features~\cite{Su2020}, which capture the harmonic relationships within an audio signal, are computed based on chroma features. The process is illustrated as:

\begin{equation}
T = U C , 
\end{equation}

where $C$ is the chroma feature vector and $U$ is a transformation matrix 
that maps chroma into a six-dimensional tonal space.

\subsubsection{Gammatone Frequency Cepstral Coefficients (GTCC)} 
GTCCs are similar to MFCCs but use a Gammatone filter bank instead of Mel filters~\cite{sharma2019environment}. GTCCs are more robust to noise and transient sounds compared to MFCCs. The computation of GTCCs is described as: 

\begin{equation}
C_n = \sum_{m=0}^{M-1} \log\!\left(S_g(m)\right)
\cos\!\left[ n\left(m - 0.5\right)\frac{\pi}{M} \right],
\end{equation}

where $C_n$ is the $n$-th GTCC coefficient, which represents a transformed  version of the Gammatone-filtered log spectrum. $M$ is the total number of  Gammatone filterbanks, typically chosen based on how fine-grained the frequency  analysis should be. $S_g(m)$ is the Gammatone power spectrum at the $m$-th  frequency band, computed by applying a Gammatone filterbank to the signal's power spectrum. $n$ is the index of the cepstral coefficient; it controls which coefficient is being computed in the Discrete Cosine Transform (DCT). 
$m$ is the index of the Gammatone filter; it corresponds to different frequency bands in the Gammatone filterbank. $\log(S_g(m))$ is the logarithm of the Gammatone power spectrum at the $m$-th band. The logarithm is applied to mimic human loudness perception, since the human ear perceives loudness logarithmically. 

\subsection{Stacked Features} 
We investigated stacked feature inputs to leverage complementary spectral, cepstral, tonal, and auditory characteristics to enhance CNN-based ESC performance. Inspired by the effectiveness of feature aggregation shown in prior research~\cite{gupta2024esc}, we selected several combinations to provide diverse acoustic perspectives. They are presented in Table~\ref{tab:feature-configurations}. 

After extracting individual features, as described in Section \ref{Features}, we aggregate them into a unified representation to serve as input to the CNN models. This process, known as feature stacking, involves combining multiple spectrogrambased features into a multi-channel image-like structure. Each extracted feature—such as MFCC, GTCC, Log-Mel Spectrogram, Chroma, Tonnetz, or Spectral Contrast—is initially represented as a two-dimensional array corresponding to the time-frequency domain. However, the shape of these arrays can differ across features. To make them compatible for stacking, we first resize all feature maps to a common dimension of 128 × 128. This resizing is achieved either through zero-padding or interpolation, depending on the original shape of the feature. Once all features share the same spatial dimensions, they are concatenated along a new axis representing channels, similar to how RGB channels are
arranged in color images. 

For instance, when combining MFCC, GTCC, and Log-Mel Spectrogram, the resulting input tensor has a shape of 128 × 128 × 3, with each “channel” corresponding to one of the three features. In the case of MFCC + GTCC + CH + LM, the resulting stacked input has a shape of 128 × 128 × 4. This approach enables the model to learn from multiple acoustic perspectives simultaneously, capturing spectral, cepstral, and tonal characteristics of the sound. Additionally, the
image-like structure of the input makes it naturally suited for convolutional neural networks, which excel at processing spatially correlated data. As shown in our experiments, this method of aggregation is both computationally efficient and effective for improving classification performance.

%\begin{table}[t]
%\begin{table*}[t]
%\begin{table}[H]
\begin{table*}[htbp]
\centering
\caption{Feature Configurations and their Descriptions.}
% \begin{tabular}{|l|p{9cm}|}
\bgroup
\def\arraystretch{1.2}%  1 is the default, change whatever we need
\begin{tabular}{l p{9cm}}
% \hline
\toprule
\textbf{Feature Configuration} & \textbf{Expected Contribution} \\ 
% \hline
\midrule
LM & Serves as a baseline, capturing spectral energy distribution. \\ 
\hline
LM + TZ & Combines spectral and tonal features, improving discrimination between harmonic and non-harmonic sounds. \\ 
\hline
LM + SPC + CH & Emphasizes distinctions between harmonic-rich, percussive, and transient sounds through combined spectral contrast and pitch class energy. \\ 
\hline
MFCC + LM & Integrates cepstral and spectral information, enhancing classification across environmental sounds. \\ 
\hline
MFCC + TZ & Enhances the classification of structured (tonal) and unstructured (non-tonal) sounds, such as distinguishing musical notes from random noise. \\ 
\hline
MFCC + GTCC + CH + LM & Provides the most comprehensive representation, integrating cepstral, spectral, tonal, and auditory system characteristics, thus improving robustness to noise. \\ 
% \hline
\bottomrule
\end{tabular}
\egroup
\label{tab:feature-configurations}
\end{table*}

\subsection{Deep Learning Architectures} 
In this work, we employed two CNNs and an AST model, which are explained in detail below.

\subsubsection{Convolutional Neural Network (CNN) architectures} 
We explore two CNN architectures designed for environmental sound classification. Both architectures are tailored to process multi-channel feature representations derived from stacked spectral, cepstral, and tonal features, enabling them to learn discriminative patterns in environmental sounds.  

\begin{figure*}
    \centering
    \includegraphics[width=0.7\linewidth]
    %\includesvg[width=\textwidth]
    {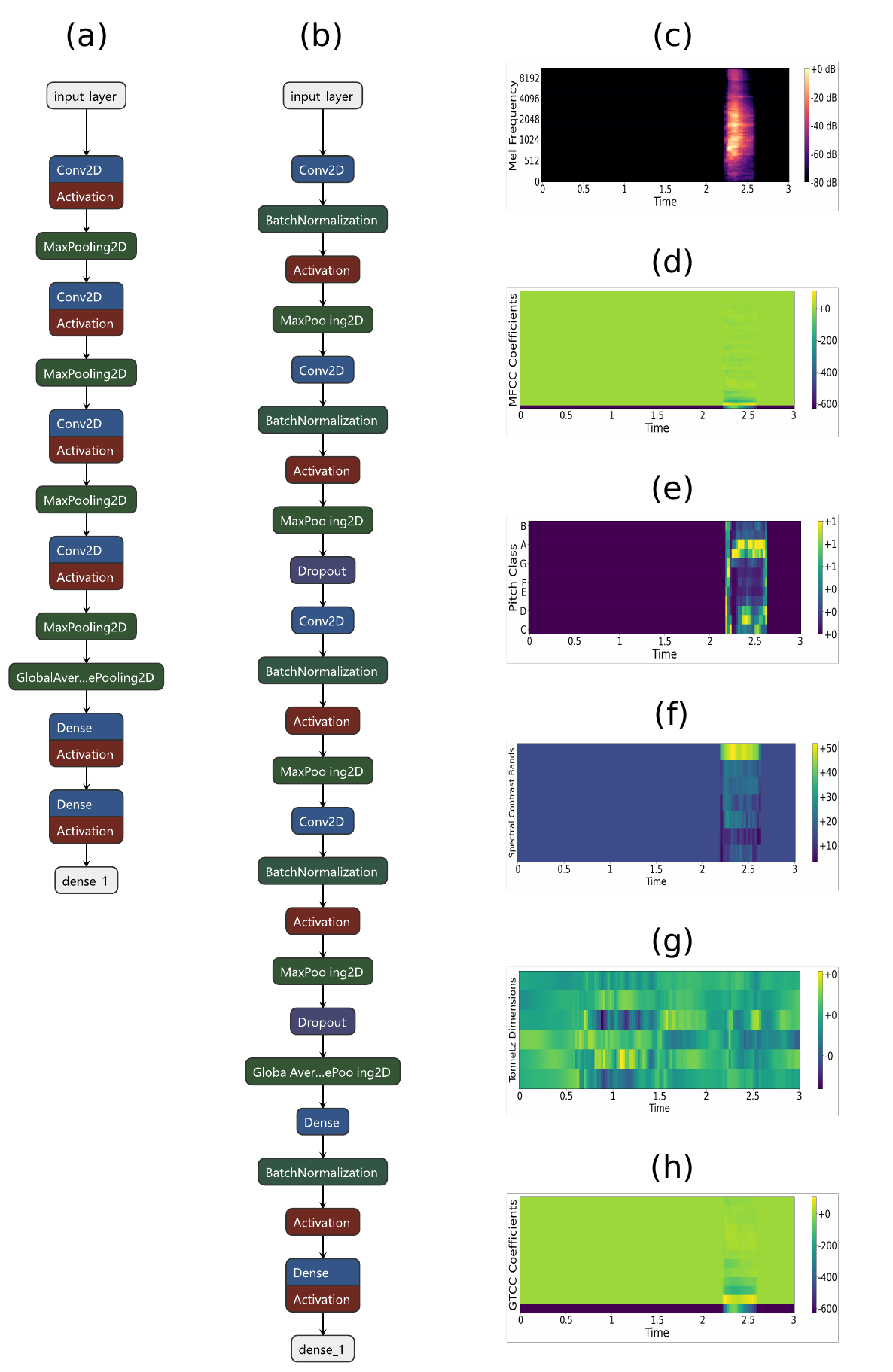}
    \caption{CNN Architectures and examples of feature applied to a sample ("dog" sample from ESC-50): (a) CNN-1 architecture. (b) CNN-2 architecture. (c) Log-Mel spectrogram. (d) MFCC. (e) Chroma. (f) Spectral contrast. (g) Tonnetz. (h) GTCC.}
    \label{fig:cnn_architectures_and_features}
\end{figure*}

\paragraph{CNN Model 1 (Baseline CNN):} The first model (CNN-1), as we have shown in Figure\ref{fig:cnn_architectures_and_features}-a, is a straightforward yet effective CNN architecture consisting of four convolutional layers, followed by global average pooling and dense layers.
This design is intended to efficiently extract hierarchical features while maintaining a balance between model complexity and computational efficiency. The architecture comprises four convolutional layers with 2×2 kernels, utilizing the ReLU
activation function. The feature extraction starts with 32 filters in the two initial layers and progresses to 64 filters for
the other two layers, enabling the network to learn increasingly complex representations. Each convolutional layer is
followed by max pooling, which reduces the spatial dimensions of the feature maps, minimizing computational cost while preserving essential patterns. To further refine the extracted features, a global average pooling layer aggregates spatial features into a single feature vector, reducing dimensionality before being fed into the dense layers. The final classification is performed using a 1024-unit dense layer with ReLU activation, followed by a softmax layer for multi-class sound recognition. The model is compiled using the Adam optimizer with a learning rate of 0.001 and is
trained using the categorical cross-entropy loss function to optimize performance in multi-class classification. 

\paragraph{CNN Model 2 (Enhanced CNN with Batch Normalization and Dropout):} The second model (CNN-2) as we have shown in Figure\ref{fig:cnn_architectures_and_features}-b, is an enhanced CNN architecture that incorporates batch normalization and dropout to improve training stability and generalization. While it follows a similar structure to CNN-1, it introduces additional regularization techniques to mitigate overfitting and optimize learning efficiency. In this architecture, Batch Normalization is applied after each convolutional and dense layer to normalize activations, improve gradient flow, and accelerate training convergence. Dropout (0.25 probability) was incorporated after the second and fourth convolutional layers to reduce overfitting by preventing co-adaptation of neurons. Global Average Pooling is utilized to reduce feature dimensionality while retaining essential spatial information before the dense layers. A 1024-unit dense layer with batch normalization and ReLU activation is followed by a softmax classifier for multi-class sound recognition. 

The selection of architectural parameters for both CNN-1 and CNN-2 was guided by a combination of empirical tuning and insights from prior research. Initially, we experimented with various CNN configurations using different numbers of convolutional layers, kernel sizes, and filter counts. Through systematic testing on the ESC-50 dataset, we observed that four convolutional layers with small kernel sizes (2×2) provided a good balance between feature learning capacity and
overfitting prevention. Similarly, the use of 32 and 64 filters in early and deeper layers, respectively, was found to yield
strong performance while keeping the model computationally efficient. The choice of global average pooling and dense
layers was inspired by common best practices in audio classification tasks and further validated through experimentation.
The dropout rate (0.25) and use of batch normalization in CNN-2 were adopted from standard regularization strategies as explored in \cite{Mushtaq2020}, \cite{hershey2017cnn}, and similar audio classification works. We also tested different dropout rates (0.3, 0.5) and optimizer learning rates (e.g., 0.0001, 0.0005, 0.001), ultimately selecting those that led to stable convergence and the best validation performance. Thus, the final CNN configurations were the result of a combination of design inspiration from prior works and hyperparameter tuning on our development set. This process ensured that the models were both
effective and generalizable to the target classification tasks.  

\subsubsection{Audio Spectrogram Transformer (AST) architecture} 
The AST employed in this work follows the base configuration introduced by the pioneer AST model~\cite{gong2021ast}. It adapts the vision transformer (ViT) framework to log-Mel spectrogram inputs, enabling long-range temporal-frequency modeling through multi-head self-attention. 
Each audio waveform is first converted into a 128-bin log-Mel spectrogram at a 16 kHz sampling rate. The spectrogram is then partitioned into a non-overlapping 16 x 16 time-frequency patches, each one linearly projected into a 768-dimensional embedding space via a convolutional patch-embedding layer. A learnable classification token (CLS) is attached to the sequence, and two-dimensional sinusoidal positional embeddings are added to retain spatial information.
The resulting sequence passes through 20 transformer encoder blocks, each composed of a multi-head self-attention module with 20 heads and a feed-forward network of 3072 hidden units with GELU activations. Residual connections, layer normalization, and dropout (rate = 0.1) are applied throughout to enhance generalization. 

For classification, the final hidden state corresponding to the CLS token is normalized and fed into a linear layer projecting to the 50 classes of the ESC-50 dataset. In the case of fine-tuning, the network is initialized from an Audioset pretrained checkpoint, enabling transfer of acoustic representations of the Audioset dataset. Optimization uses the AdamW optimizer with weight decay of $10^{-4}$, a cosine learning-rate schedule starting at $5 \times 10^{-5}$, and a batch size of 8. Training proceeds for 10 epochs. 

\subsubsection{Retraining the models using UrbanSound8K dataset} 
To assess the generalization capability of the pre-trained model, we fine-tune it using the UrbanSound8K dataset, a widely used benchmark for environmental sound classification. This retraining process is designed to adapt the ESC-50 pre-trained model to a new dataset while preserving its previously learned feature representations, thus improving its performance on different types of audio data. Instead of training the model from scratch, we utilize transfer learning.
This involves leveraging the knowledge embedded in the ESC-50 model and modifying its classification head to align with the specific label structure of UrbanSound8K. By doing so, we efficiently adapt the model to a new domain while retaining valuable learned features from the original training process.

The input structure of the model remains unchanged, with retraining applied exclusively to the final classification layer.
This selective fine-tuning approach ensures that the lower-level feature extraction layers, which have already learned
meaningful audio representations, remain intact, allowing the model to generalize well across different datasets. To achieve optimal adaptation, the model is trained for 50 epochs with a batch size of 32, striking a balance between stability and convergence speed. This training strategy enables the model to refine its decision boundaries for the new dataset while preventing overfitting and ensuring that previously learned knowledge is effectively transferred.

\subsection{Experimental Setup} 
To evaluate the performance of our proposed approach for environmental sound classification, we conducted a series
of experiments using two CNN architectures (CNN-1 and CNN-2) on the UrbanSound8K and ESC-50 datasets. The UrbanSound8K dataset contains 8,732 labeled audio clips (up to 4 seconds each) spanning 10 environmental sound classes (e.g., car horn, siren, dog bark), and follows a 10-fold cross-validation structure. The ESC-50 dataset consists of 2,000 five-second clips organized into 50 balanced classes. 

Our experimental pipeline involves four main stages: data preprocessing, feature extraction, model training (pre-training
and fine-tuning), and performance evaluation. All audio clips were resampled to 22,050 Hz and trimmed or zeropadded to 3 seconds. Feature values were normalized to zero mean and unit variance. Each clip was transformed into multiple time-frequency representations, including MFCC, GTCC, Chroma, Tonnetz, Spectral Contrast, and Log-Mel
Spectrogram. These features were resized to a uniform dimension of 128 × 128 using zero-padding or interpolation, and then stacked to form multi-channel inputs analogous to RGB images. For training, each CNN model was initially pre-trained on ESC-50 and fine-tuned on UrbanSound8K using transfer learning. During fine-tuning, pre-trained
convolutional layers were frozen and only the final classification layers were retrained. The models were trained using
categorical cross-entropy loss and the Adam optimizer, with accuracy as the main evaluation metric. Training was conducted for 150 epochs on ESC-50 and 50 epochs on UrbanSound8K, using a batch size of 32. Validation was performed using held-out test sets to evaluate generalization. 

All experiments were carried out on Google Colab, leveraging its NVIDIA H100 GPU for accelerated computation. The models were implemented using the Keras API with TensorFlow as the backend, and Librosa was employed for audio signal processing and feature extraction. To further assess the effect of different feature combinations, we conducted 5-fold cross-validation experiments using both CNN architectures on the ESC-50 dataset. The results, presented in Table~\ref{tab:cnn_results}, report classification accuracy across each fold along with the average performance.

\begin{table*}[ht]
%\begin{table}[H]
%\begin{table*}[h]
\centering
\caption{Accuracy of CNN-1 and CNN-2 across 5 folds with different feature combinations.}
\label{tab:cnn_results}
\begin{tabular}{c c c c c c c c}
\hline
\textbf{Models} & \textbf{Features} & \textbf{Fold 1} & \textbf{Fold 2} & \textbf{Fold 3} & \textbf{Fold 4} & \textbf{Fold 5} & \textbf{Average Accuracy} \\
\hline
\multirow{6}{*}{CNN-1} & LM & 0.70 & 0.67 & 0.69 & 0.66 & 0.66 & \textbf{0.68} \\
 & LM+TZ & 0.66 & 0.66 & 0.68 & 0.62 & 0.63 & 0.65 \\
 & LM+MFCC & 0.65 & 0.64 & 0.66 & 0.64 & 0.61 & 0.64 \\
 & MFCC+TZ & 0.65 & 0.59 & 0.61 & 0.60 & 0.64 & 0.62 \\
 & LM+SPC+CH & 0.61 & 0.59 & 0.63 & 0.62 & 0.65 & 0.62 \\
 & MFCC+GTCC+CH+LM & 0.69 & 0.67 & 0.65 & 0.66 & 0.68 & 0.67 \\
\hline
\multirow{6}{*}{CNN-2} & LM & 0.43 & 0.47 & 0.45 & 0.46 & 0.44 & 0.45 \\
 & LM+TZ & 0.65 & 0.62 & 0.69 & 0.68 & 0.66 & \textbf{0.66} \\
 & LM+MFCC & 0.61 & 0.62 & 0.57 & 0.60 & 0.58 & 0.59 \\
 & MFCC+TZ & 0.62 & 0.59 & 0.63 & 0.62 & 0.65 & 0.62 \\
 & LM+SPC+CH & 0.52 & 0.56 & 0.54 & 0.51 & 0.53 & 0.53 \\
 & MFCC+GTCC+CH+LM & 0.59 & 0.56 & 0.59 & 0.57 & 0.60 & 0.58 \\
\hline
\end{tabular}
\end{table*}

\section{Results and Analysis} 
\begin{figure*}
    \centering
    \includegraphics[width=1\linewidth]{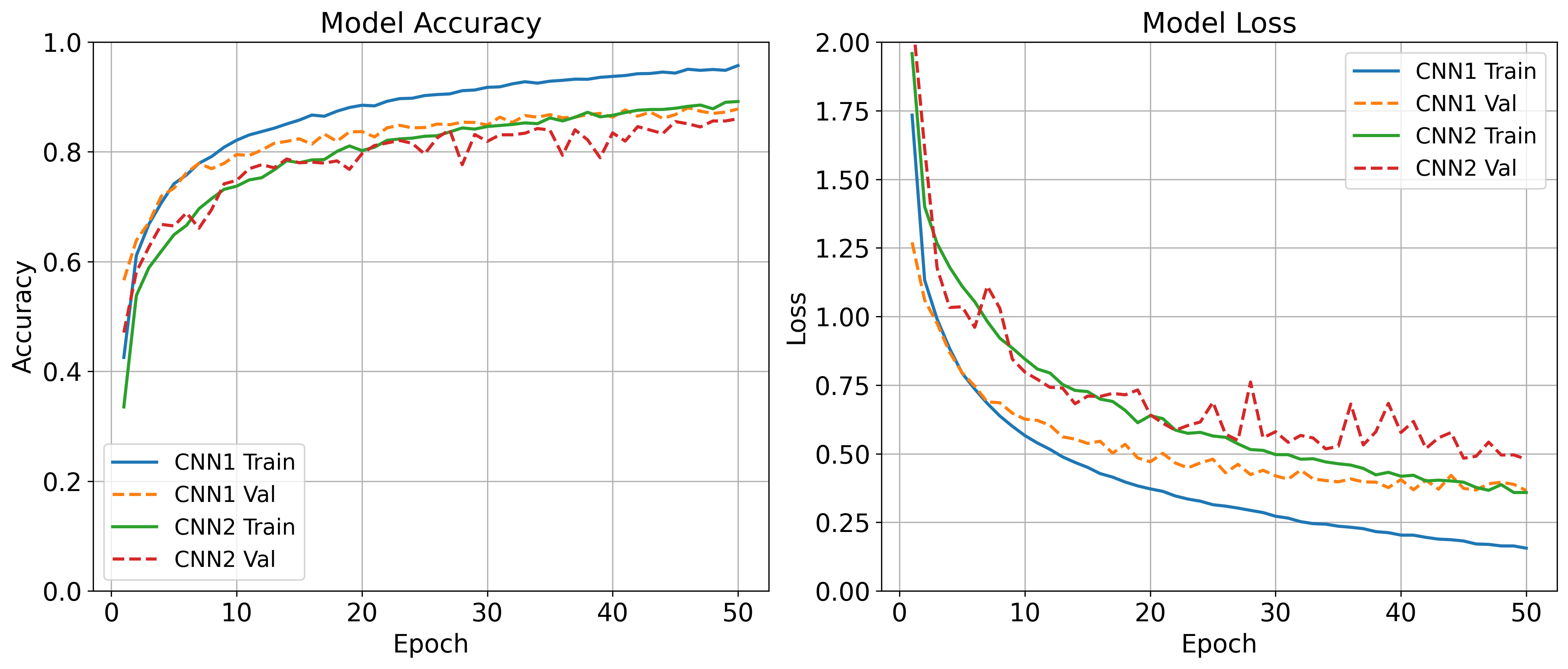}
    \caption{Comparison of Training and Validation Accuracy and Loss for CNN-1 and CNN-2 using Log-Mel (LM) Features Only.}
    \label{fig:cnn_comparison_LM_accuracy_loss}
\end{figure*}

\begin{figure*}
    \centering
    \includegraphics[width=1\linewidth]{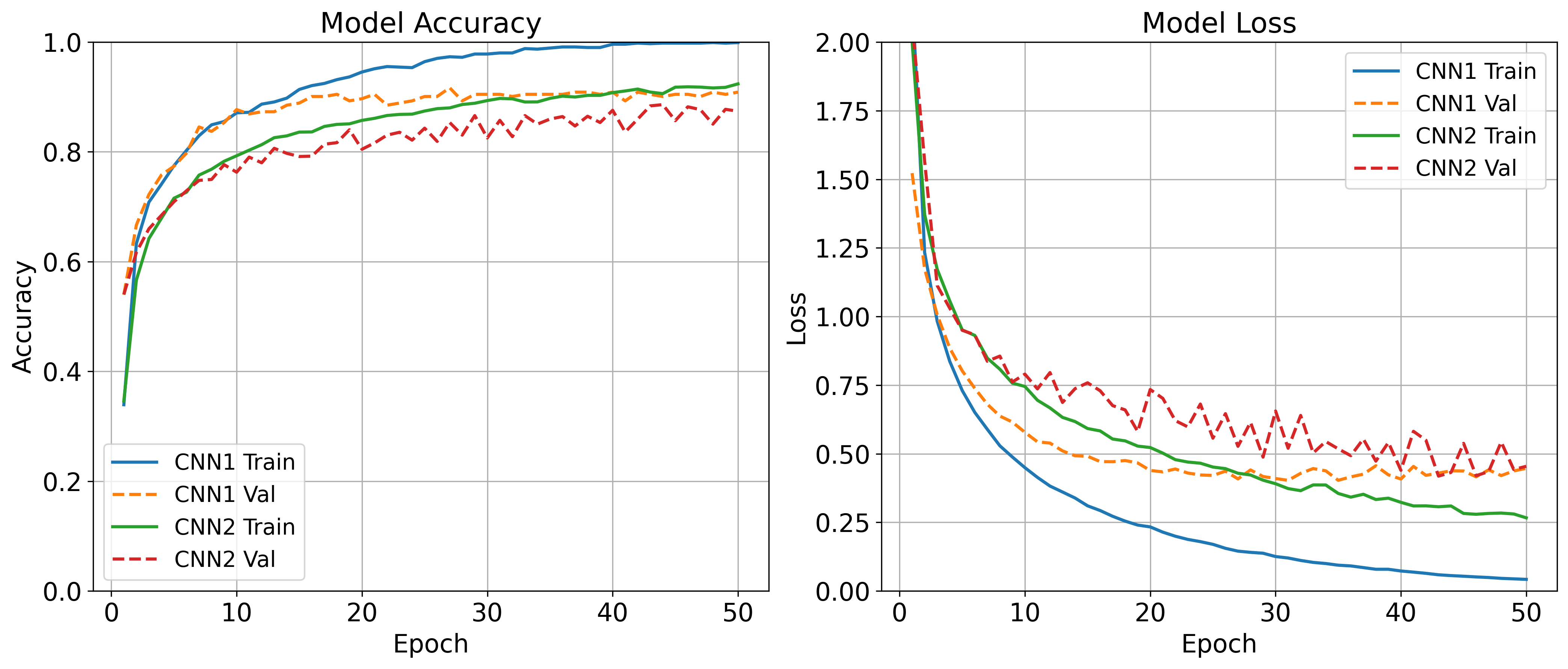}
    \caption{Comparison of Training and Validation Accuracy and Loss for CNN-1 and CNN-2 using stacked Log-Mel (LM) and MFCC features.}
    \label{fig:cnn_comparison_LM_MFCC_accuracy_loss}
\end{figure*}

\begin{figure*}
    \centering
    \includegraphics[width=1\linewidth]{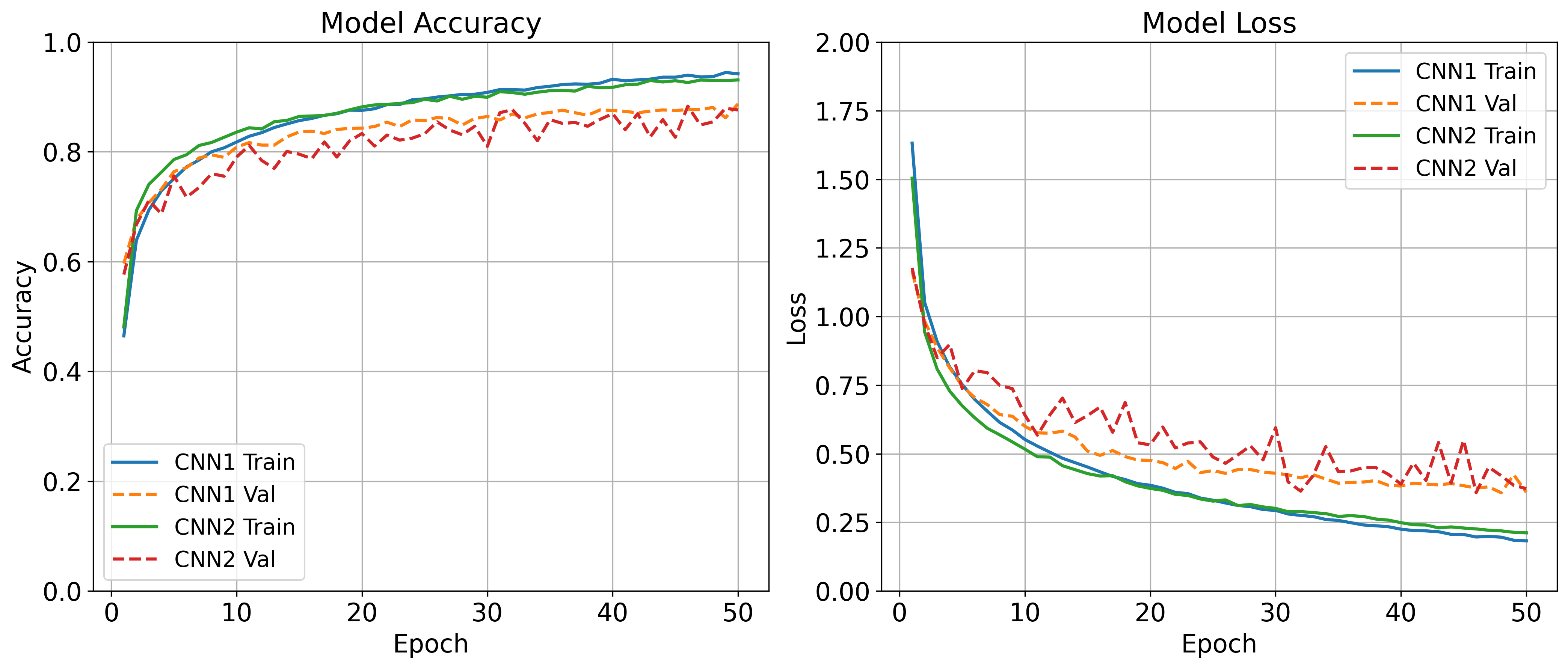}
    \caption{Comparison of Training and Validation Accuracy and Loss for CNN-1 and CNN-2 using stacked Log-Mel (LM) and Tonnetz (TZ) features.}
    \label{fig:cnn_comparison_LM_TZ_accuracy_loss}
\end{figure*}

\begin{figure*}
    \centering
    \includegraphics[width=1\linewidth]{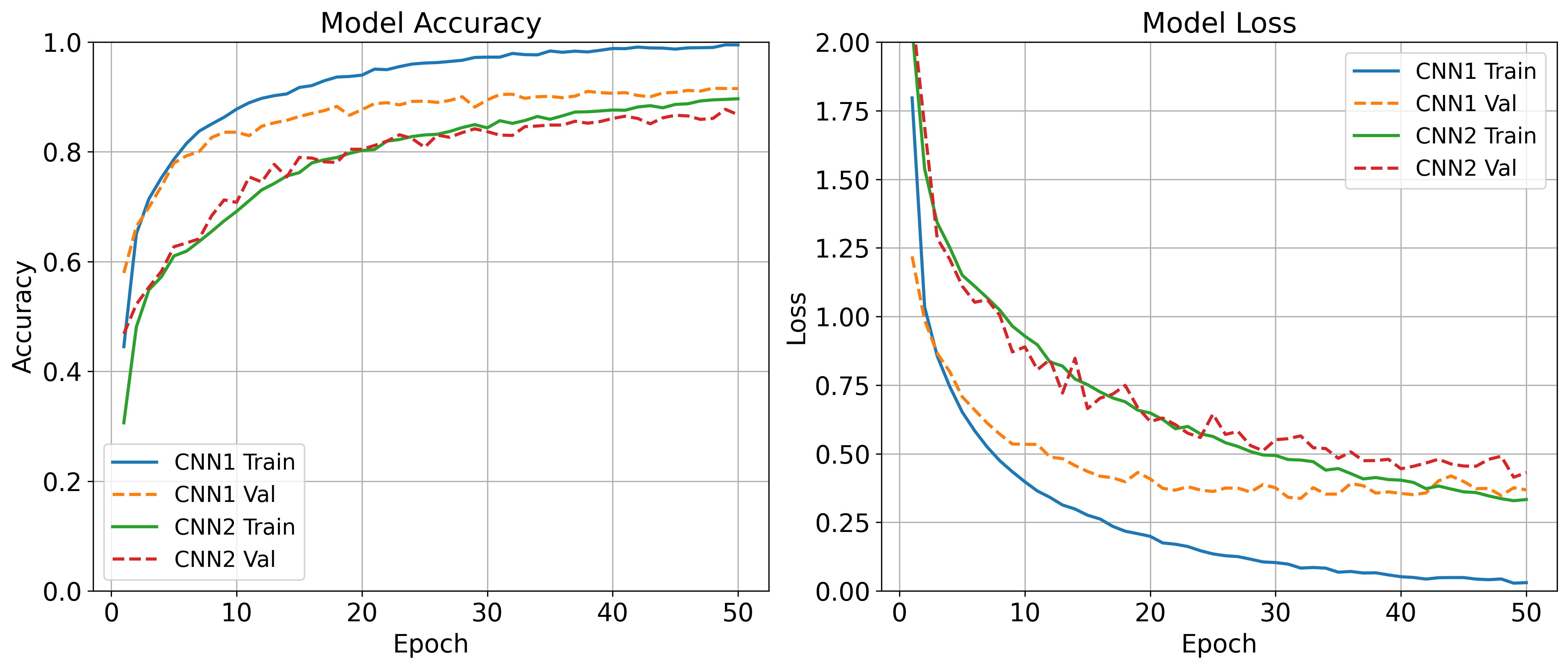}
    \caption{Comparison of Training and Validation Accuracy and Loss for CNN-1 and CNN-2 using stacked MFCC and TZ features.}
    \label{fig:cnn_comparison_MFCC_TZ_accuracy_loss}
\end{figure*}

\begin{figure*}
    \centering
    \includegraphics[width=1\linewidth]{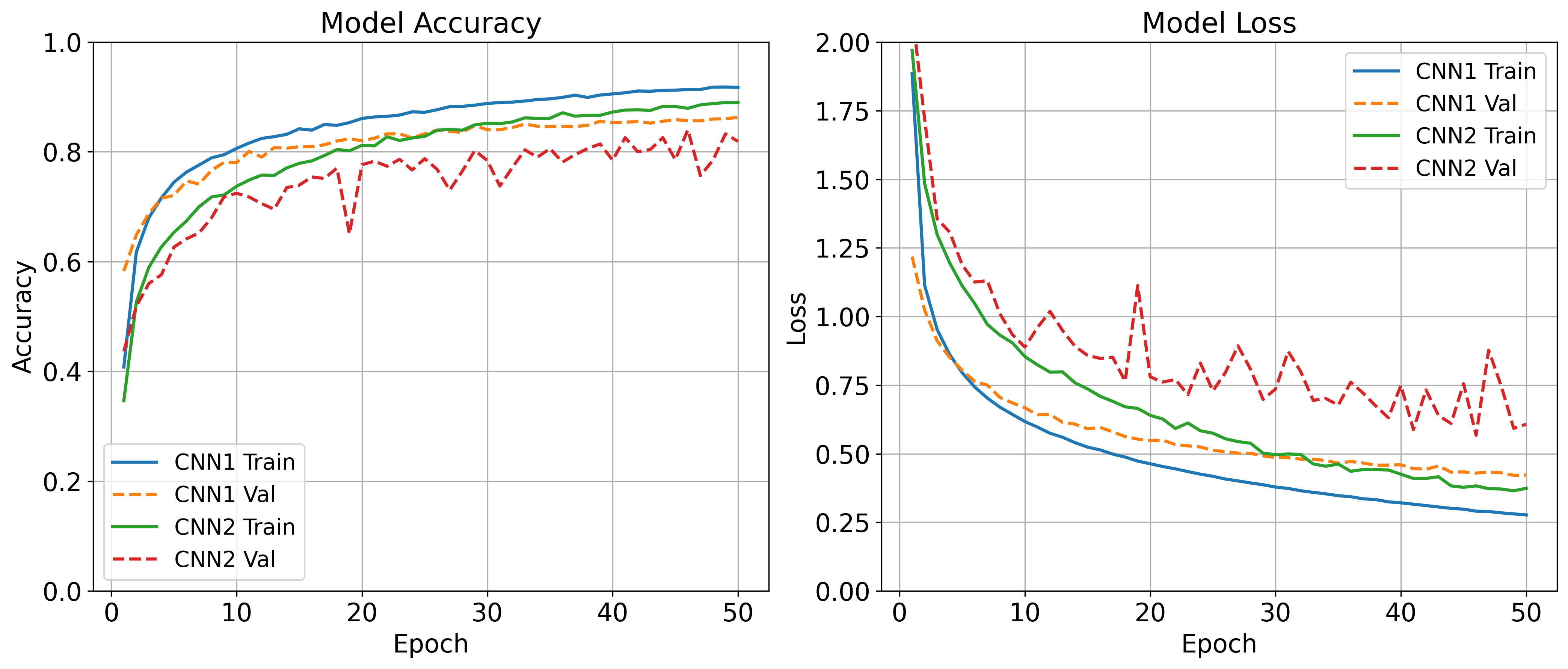}
    \caption{Comparison of Training and Validation Accuracy and Loss for CNN-1 and CNN-2 using stacked Log-Mel (LM), Spectral Contrast (SPC), and Chroma (CH) features.}
    \label{fig:cnn_comparison_LM_SPC_CH_accuracy_loss}
\end{figure*}

\begin{figure*}
    \centering
    \includegraphics[width=1\linewidth]{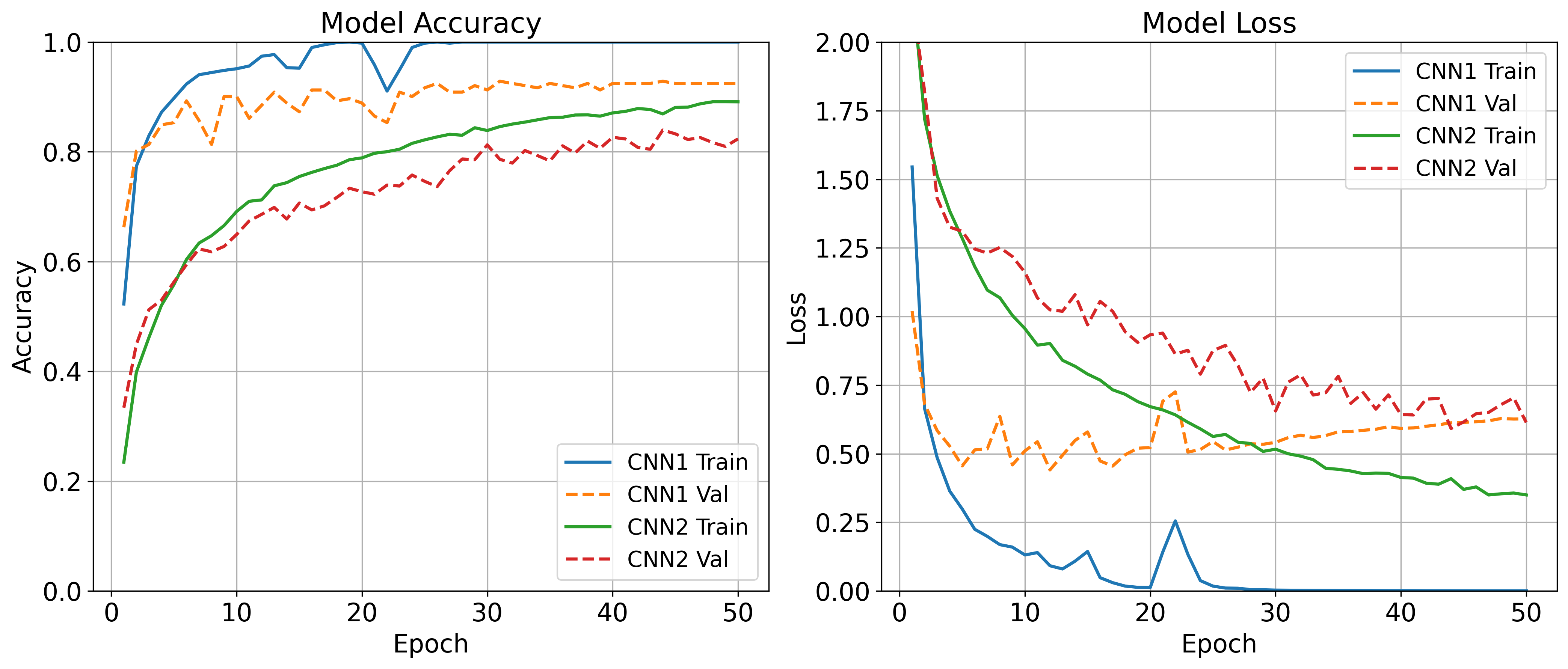}
    \caption{Comparison of training and validation accuracy and loss for CNN-1 and CNN-2 using stacked MFCC, Chroma, Mel-spectrogram, and GTCC features.}
    \label{fig:cnn_comparison_mfcc_chroma_mel_gtcc}
\end{figure*}

Figure \ref{fig:cnn_comparison_LM_accuracy_loss} to Figure \ref{fig:cnn_comparison_mfcc_chroma_mel_gtcc} illustrates the training and validation accuracy and loss of the CNN-1 and CNN-2 models during
fine-tuning on the UrbanSound8K dataset, following pre-training on ESC-50, when different input features are used. 

\begin{table*}[htbp] 
\centering
\small
\setlength{\tabcolsep}{4pt}
\caption{Performance comparison of CNN models}
\begin{tabular}{c p{3.2cm} l c rrrrrr}
\toprule
Model & Features & Training Setup & B.N & Val.Acc & Train.Acc & Precision & Recall & F1-score & Epochs \\
\midrule
1 & LM & ESC, All.L & No & 0.68 & 1.00 & 0.68 & 0.68 & 0.66 & 150 \\
1 & LM+TZ & ESC, All.L & No & 0.65 & 1.00 & 0.66 & 0.66 & 0.64 & 150 \\
1 & LM+MFCC & ESC, All.L & No & 0.64 & 1.00 & 0.68 & 0.64 & 0.63 & 150 \\
1 & MFCC+TZ & ESC, All.L & No & 0.62 & 1.00 & 0.65 & 0.62 & 0.61 & 150 \\
1 & LM+SPC+CH & ESC, All.L & No & 0.62 & 1.00 & 0.65 & 0.62 & 0.62 & 150 \\
1 & MFCC+GTCC+CH+LM & ESC, All.L & No & 0.67 & 1.00 & 0.70 & 0.67 & 0.67 & 150 \\
2 & LM & ESC, All.L & Yes & 0.45 & 0.68 & 0.59 & 0.45 & 0.44 & 150 \\
2 & LM+TZ & ESC, All.L & Yes & 0.66 & 0.98 & 0.71 & 0.67 & 0.65 & 150 \\
2 & LM+MFCC & ESC, All.L & Yes & 0.59 & 0.99 & 0.64 & 0.59 & 0.59 & 150 \\
2 & MFCC+TZ & ESC, All.L & Yes & 0.62 & 0.97 & 0.69 & 0.63 & 0.61 & 150 \\
2 & LM+SPC+CH & ESC, All.L & Yes & 0.53 & 0.81 & 0.67 & 0.54 & 0.54 & 150 \\
2 & MFCC+GTCC+CH+LM & ESC, All.L & Yes & 0.58 & 0.97 & 0.67 & 0.59 & 0.56 & 150 \\
1 & LM & ESC+US8K, Last.L & No & 0.87 & 0.95 & 0.88 & 0.88 & 0.87 & 50 \\
1 & LM+TZ & ESC+US8K, Last.L & No & 0.88 & 0.96 & 0.89 & 0.88 & 0.88 & 50 \\
1 & LM+MFCC & ESC+US8K, Last.L & No & 0.91 & 0.98 & 0.92 & 0.92 & 0.92 & 50 \\
1 & MFCC+TZ & ESC+US8K, Last.L & No & 0.91 & 0.99 & 0.92 & 0.91 & 0.92 & 50 \\
1 & LM+SPC+CH & ESC+US8K, Last.L & No & 0.85 & 0.92 & 0.86 & 0.85 & 0.85 & 50 \\
1 & MFCC+GTCC+CH+LM & ESC+US8K, Last.L & No & \textbf{0.92} & 1.00 & 0.92 & 0.92 & 0.92 & 50 \\
2 & LM & ESC+US8K, Last.L & Yes & 0.85 & 0.91 & 0.86 & 0.85 & 0.85 & 50 \\
2 & LM+TZ & ESC+US8K, Last.L & Yes & 0.85 & 0.89 & 0.85 & 0.85 & 0.85 & 50 \\
2 & LM+MFCC & ESC+US8K, Last.L & Yes & 0.86 & 0.90 & 0.87 & 0.86 & 0.86 & 50 \\
2 & MFCC+TZ & ESC+US8K, Last.L & Yes & 0.87 & 0.92 & 0.87 & 0.87 & 0.87 & 50 \\
2 & LM+SPC+CH & ESC+US8K, Last.L & Yes & 0.85 & 0.89 & 0.86 & 0.85 & 0.85 & 50 \\
2 & MFCC+GTCC+CH+LM & ESC+US8K, Last.L & Yes & 0.87 & 0.90 & 0.88 & 0.87 & 0.87 & 50 \\
\bottomrule
\end{tabular}
\label{tab:cnn_summary}
\end{table*}

%%%%%%%%%%%%%%%%%%%%%%%%%%% 

%%%%%%%%%%%%%%%%%%%%%%%% 

%%%%%%%%%%%%%%%%%%%%%%%% 

%%%%%%%%%%%%%%%%%%%%%%%%%% 

%%%%%%%%%%%%%%%%%%%%%%%%%%%% 

%%%%%%%%%%%%%%%%%%%%%%%%%555

\subsection{Performance Trends and Observations} 
In this subsection, we provide a comprehensive analysis of the experimental findings, focusing on the performance behavior of the proposed models under various conditions. As shown in Table \ref{tab:cnn_summary}, the column "Training Setup" indicates the transfer learning strategy used during training. “All.L” refers to training all convolutional layers from scratch (i.e., no layers are frozen) and “Last.L” refers to fine-tuning only the last layer of the network while freezing the rest, which allows for transfer learning from pretrained weights on the ESC-50 dataset to the UrbanSound8K dataset. A significant performance improvement is observed when moving from the ESC, All.L setup to the ESC + US8K, Last.L setup. This improvement is primarily due to the transfer learning strategy and the increased diversity of training data rather than architectural modifications alone. In the ESC, All.L configuration, the networks are trained from scratch using only ESC-50 data, which limits the amount of available acoustic variability. In contrast, the ESC + US8K, Last.L configuration leverages pretrained representations learned from ESC-50 and adapts them to UrbanSound8K through fine-tuning of the final layer. This cross-dataset transfer exposes the models to a broader distribution of environmental sounds, enabling more robust feature representations and better generalization. Therefore, the observed accuracy gains are largely attributable to the expanded data regime and transfer-learning strategy rather than solely to architectural differences.

Across the experiments, feature stacking played a significant role in improving accuracy. While single-feature models (LM alone) achieved reasonable performance with CNN-2 reaching 85.45\% validation accuracy and CNN-1 reaching 86.83\%, the fusion of MFCC, GTCC, CH, and HLM led CNN-1 to achieve 92.46\% validation accuracy, making it the best-performing configuration. CNN-2 trained on the same multi-feature combination achieved 86.71\%, confirming that combining complementary feature sets enhances overall classification performance.

When comparing model architectures, CNN-1 consistently outperformed CNN-2 across all feature sets. Although CNN-2 exhibited slightly lower recall, indicating higher misclassification rates for certain classes, the overall accuracy and stability of CNN-1 remained superior. Batch normalization was applied in CNN-2 but not in CNN-1; however, CNN-1 still outperformed CNN-2, suggesting that its architecture is inherently more effective at learning transferable representations, even without normalization layers. CNN-2, on the other hand, benefited from batch normalization in terms of smoother convergence, though with slightly lower validation accuracy.

Analyzing precision, recall, and F1-score reveals that the best-performing model—CNN-1 trained on MFCC+GTCC +CH+LM achieved 100\% training accuracy, 92.34\% precision, and 92.46\% recall, demonstrating strong generalization to unseen samples. CNN-2 models showed slightly lower recall, indicating more difficulty in identifying minority sound classes, though precision remained high.

\subsection{Comparison with AST Model} 
The comparative evaluation between the CNN and the AST models reveals a clear performance distinction across different training configurations, as demonstrated in Table~\ref{tab:accuracy}. When trained on the ESC-50 dataset from scratch (All.L), the AST model using a single Mel-Spectrogram input achieved a validation accuracy of 43\%, lower than the results obtained by both CNN models 1 and 2, which were 68\% (using LM) and 66\% (using LM+TZ), respectively. The
performance gap underscores the AST model’s sensitivity to limited training data, as its self-attention mechanism typically requires larger datasets or extensive pretraining to reach higher accuracy ~\cite{gong2021ast}. 

When employing transfer learning by fine-tuning only the last layer on UrbanSound8K after ESC-50 pretrained(Last.L),the AST model improved to 58\% validation accuracy. While this represents a gain compared to its trained-from-scratch result, it remains lower than the CNN models. These results indicate that, even under transfer learning conditions, using ESC-50 and UrbanSound8K, CNNs leverage stacked features with an advantage over the AST. 

\begin{table*}[htbp]
\centering
\small
\setlength{\tabcolsep}{6pt}
\caption{Validation accuracy for CNN architectures and the AST model.}
\begin{tabular}{l p{5cm} l c}
\toprule
\textbf{Model} & \textbf{Features} & \textbf{Training Setup} & \textbf{Val. Accuracy} \\
\midrule

CNN-1 & LM & ESC, All.L & \textbf{0.68} \\
CNN-2 & LM + TZ & ESC, All.L & 0.66 \\
AST & Mel Spectrogram & ESC, All.L & 0.43 \\

CNN-1 & MFCC + GTCC + CH + LM & ESC + US8K, Last.L & \textbf{0.92} \\
CNN-2 & MFCC + GTCC + CH + LM & ESC + US8K, Last.L & 0.87 \\

AST & Mel Spectrogram & ESC + US8K, Last.L & 0.58 \\
AST & Mel Spectrogram & Audioset + ESC, Last.L & \textbf{0.99} \\

\bottomrule
\end{tabular}
\label{tab:accuracy}
\end{table*}

The performance difference can be attributed to both architectural and data factors. CNNs, particularly when supplied with stacked features, effectively capture complementary time-frequency representations and additional representations when aggregated with other features, enabling them to generalize well even with moderate training data sizes. The AST, while capable of modeling long-range dependencies, is more reliant on a large-scale pretraining dataset to fully exploit its attention mechanisms. We evaluated the AST with Mel-Spectrogram input using AudioSet pretraining followed by
ESC fine-tuning, which achieved a validation accuracy of 99\%, confirming the boost that large-scale pretraining can provide. However, without the large-scale pretraining, AST struggles to match the capabilities of CNNs.

\subsection{Training Time and Computational Efficiency Analysis} 
The computational efficiency of a deep learning model is a key factor in real-world applications, particularly for
deployment on resource-constrained or real-time systems. In our experiments, we analyzed the training time and inference efficiency of the CNN-1 and CNN-2 models when fine-tuned on the UrbanSound8K dataset using a last-layer retraining strategy.

As shown in Table \ref{tab:time}, CNN-1 has fewer parameters (146K) compared to CNN-2 (151K), reflecting its slightly simpler architecture. In terms of inference efficiency, CNN-1 achieved a mean inference time of 21.92 ms per sample with a standard deviation of 2.15 ms, while CNN-2 exhibited a higher inference cost, with a mean inference time of 30.95 ms and a standard deviation of 4.07 ms. These results were obtained over 50 inference iterations per sample to ensure statistical reliability. This difference is primarily attributed to the inclusion of Batch Normalization layers in CNN-2, which introduce additional computations during inference despite their benefits for training stability.
Regarding training efficiency, CNN-1 consistently converged faster than CNN-2, requiring fewer training iterations to reach its best validation performance. The simpler architecture of CNN-1 enables more efficient gradient updates and lower computational overhead during training. In contrast, CNN-2 demonstrated improved training stability and smoother convergence behavior, benefiting from the regularization effect of Batch Normalization, particularly in later training stages.

Considering both training and inference efficiency, CNN-1 is better suited for real-time or embedded applications where low latency and reduced computational cost are critical. CNN-2, while computationally more demanding, may be preferred in scenarios where training stability and marginal performance improvements are prioritized and computational resources are sufficient.
When compared with the AST model, a substantial difference in computational complexity is observed. Although AST offers competitive classification performance, its transformer-based architecture involves self-attention mechanisms and a significantly larger parameter set (86M parameters), resulting in a much higher inference time of 1.1 s per sample. This makes AST less suitable for real-time or resource-limited environments. Therefore, depending on the application requirements, a trade-off between inference speed and classification accuracy must be carefully considered.

\begin{table}[h!]
\caption{Comparison of models in terms of parameters and inference time.}
\centering
\label{tab:time}
\begin{tabular}{l>{\centering\arraybackslash}m{2cm}>{\centering\arraybackslash}m{3.5cm}}%>{\centering\arraybackslash}m{2cm}}
\toprule
\textbf{Model} & \textbf{Number of Parameters} & \textbf{Inference Time per Sample} \\%& \textbf{Note} \\
\midrule
CNN-1 & 146 K & \textbf{21.92 ms} \\ 
CNN-2 & 151 K & 30.95 ms  \\ 
AST  & 86 M     & 1.1 s     \\ 
\bottomrule
\end{tabular}
\end{table}  

\subsection{Discussion on Generalization and Overfitting} 
CNN-1 demonstrates a more balanced trade-off between generalization performance and computational efficiency compared to CNN-2. Although CNN-2 benefits from improved training stability due to Batch Normalization, this advantage comes at the cost of increased inference latency.
In terms of validation performance, CNN-1 consistently achieved higher accuracy on unseen data, indicating stronger generalization capabilities, particularly when trained with multi-feature inputs. This behavior suggests that CNN-1 effectively learns discriminative representations without relying on deeper regularization mechanisms.

Furthermore, CNN-1 exhibits superior computational efficiency both during training and inference. Its faster convergence reduces training cost, while its lower inference latency makes it more suitable for real-time deployment. In contrast, CNN-2 offers improved training stability and smoother convergence, which may be advantageous in scenarios where training robustness is more critical than inference speed.
These results highlight that CNN-1 provides a favorable balance between performance, training efficiency, and inference latency, making it a strong candidate for practical sound classification systems operating under computational constraints.

\section{Conclusion} 

We investigated the impact of stacking multiple audio features, including MFCCs, log-Mel spectrograms, Chroma, Spectral Contrast, Tonnetz, and GTCC, as inputs to CNN-based sound classification models. Experimental results on the ESC-50 and UrbanSound8K datasets demonstrate that combining complementary features consistently improves classification performance by providing richer and more discriminative representations.

Among the evaluated CNN architectures, CNN-1 achieved the most favorable balance between classification performance and computational efficiency. It exhibited the lowest inference latency, making it particularly suitable for real-time and resource-constrained applications. CNN-2, which incorporates Batch Normalization, showed improved training stability and smoother convergence but incurred a higher inference cost, highlighting a trade-off between training robustness and deployment efficiency.
The AST model demonstrated strong representation learning capabilities due to its transformer-based architecture; however, its substantially higher computational cost and inference latency limit its applicability in real-time or embedded scenarios. These results emphasize the importance of selecting model architectures based on application requirements, where lightweight CNNs are preferable for low-latency deployment, while transformer-based models may be advantageous when computational resources are abundant.

This study highlights the effectiveness of stacked audio features in enhancing sound classification performance and underscores the necessity of balancing accuracy, training stability, and inference efficiency when designing models for practical deployment.
As a future research direction, these findings motivate the exploration of online and adaptive learning strategies that can incrementally update lightweight models in dynamic acoustic environments while preserving real-time inference capabilities.

\emergencystretch=2em

\bibliographystyle{IEEEtran}
\bibliography{main}
\end{document}